\documentclass[%
 aip,
 amsmath,amssymb,
 reprint,%
]{revtex4-1}

\usepackage{graphicx}
\usepackage{dcolumn}
\usepackage{bm}

\usepackage[utf8]{inputenc}
\usepackage[T1]{fontenc}
\usepackage{mathptmx}
\usepackage{etoolbox}
\usepackage[version=4]{mhchem}
\usepackage{placeins}

\makeatletter
\def\@email#1#2{%
 \endgroup
 \patchcmd{\titleblock@produce}
  {\frontmatter@RRAPformat}
  {\frontmatter@RRAPformat{\produce@RRAP{*#1\href{mailto:#2}{#2}}}\frontmatter@RRAPformat}
  {}{}
}%
\makeatother
\begin{document}
\setlength{\parskip}{0pt}
\preprint{AIP/123-QED}

\title[Self ordering to imposed ordering of dust - a continuous spatial phase transition experiment in MDPX]{Self ordering to imposed ordering of dust - a continuous spatial phase transition experiment in MDPX}
\author{Siddharth Bachoti}%
 \email{szb0234@auburn.edu}
 \affiliation{ 
Department of Physics, Auburn University (USA)
}%

\author{Saikat Chakraborty Thakur}%
 \affiliation{ 
Department of Physics, Auburn University (USA)
}%

\author{Rahul Banka}%
\affiliation{ 
Department of Materials, University of Manchester (UK)
}%

\author{Cameron Royer}%
 \affiliation{ 
Department of Physics, Auburn University (USA)
}%

\author{Edward Thomas Jr.}%
 \affiliation{ 
Department of Physics, Auburn University (USA)
}%

\date{\today}

\begin{abstract}
Previous experiments conducted in the Magnetized Dusty Plasma eXperiment (MDPX) revealed an intriguing phenomenon first referred to as imposed ordering. This occurs when micron-sized dust particles become aligned with the geometry of a conducting mesh placed above the dust (at a distance much larger than the plasma Debye length or the ion-neutral or electron-neutral mean free paths) in the presence of a strong magnetic field perpendicular to the mesh. In this work, results of a transition experiment are presented wherein starting from a classical two-dimensional Coulomb crystal with hexagonal symmetry in an unmagnetized plasma ($B = 0\,T$), dust transitions to a state in which it flows along the geometry of a conducting mesh placed above it, mapping out the 4-fold symmetry of the boundary condition. It is hypothesized that beyond a certain magnetization, elongated electric potential structures emanating from the mesh drive the dust motion to reflect the mesh morphology, transitioning from a 6-fold self ordering to 4-fold imposed ordering. The various dust phases are quantified and a critical value of magnetic field is identified in the transition experiment indicating the onset of imposed ordering.
\end{abstract}

\maketitle

\section{Introduction}\label{section:Introduction}

Dusty plasma refers to a system containing micrometer or nanometer sized particles in a plasma. When placed in a plasma, micron-sized dust particles accumulate electrons on their surface, acquiring a net negative charge of typically 5000 to 10000 electrons owing to higher mobility of electrons compared to ions. This gives rise to a variety of dynamics due to dust-dust interactions, dust-plasma interactions, and dust response to externally applied electric and magnetic fields. Therefore, dusty plasmas exhibit a variety of phenomena such as strong coupling\cite{kumar_experimental_2025}\textsuperscript{,}\cite{thomas_plasma_1994}, crystalline phase transitions\cite{singh_square_2022}\textsuperscript{,}\cite{singh_experimental_2023} and waves\cite{thomas_observations_2007}\textsuperscript{,}\cite{williams_observations_2008}. Dusty plasmas naturally occur in systems such as comet tails\cite{horanyi_dynamics_1986}, the rings of Saturn\cite{smith_new_1982}, and the Lunar surface\cite{rennilson_surveyor_1974}. These are also relevant to the semiconductor industry, where dust is an unwanted byproduct of the plasma etching processes that make integrated circuits from silicon wafer starters\cite{selwyn_plasma_1991}. Magnetized low-temperature plasmas such as those used in this study have also been used for the growth of nanoparticles\cite{ramkorun_comparing_2024}\textsuperscript{,}\cite{ramkorun_electron_2025} and hence have relevance to material science. RF plasma diagnostics represent another area in which dust particles have proven useful, as they can act as probes of electric fields in these systems \cite{hartmann_dust_2014}\textsuperscript{,}\cite{ekanayaka_high-resolution_2025}. Dusty plasmas can also be used as macroscopic analogs to address fundamental physics questions such as liquid crystals and colloids. This work investigates the motion of dust particles in the presence of externally imposed potential structures, which can be used to model colloidal dynamics over periodic substrates\cite{reichhardt_depinning_2016} or the flow of superconducting vortices on 2D films with regular arrays of defects\cite{harada_direct_1996}. 

In our experiments, we create a dusty plasma by injecting monodisperse $7.7 \mu m$ sized spherical silica particles into a capacitively coupled plasma (CCP). These weakly ionized plasmas are typically produced using low RF powers ($<10\,W$) and fall under the regime of "low-temperature plasmas" where the electron temperatures are $2-4 \,eV$ and densities are $\sim 10^{13} - 10^{14} \,m^{-3}$.

In this study, experiments are performed in the Magnetized Dusty Plasma eXperiment (MDPX) device\cite{jr_magnetized_2015}.  The MDPX device consists a superconducting magnet with a large central bore where different plasma chambers can be inserted.  Details of the MDPX configuration used for this study are discussed in Section \ref{section:Experiment setup}.

Two separate earlier experiments were conducted in different plasma chambers, boundary conditions, magnetic fields and pressures to obtain two thermodynamically distinct dust crystal states in the MDPX device. In one set of experiments reported in Jaiswal et al.\cite{jaiswal_effect_2017}, a dust crystal was obtained at $B = 0\,T$ and $200\,mTorr$ of background neutral pressure under uniform boundary condition (a uniform, conducting fluorine doped tin oxide (FTO) glass in the chamber above the dust). Starting with these initial conditions, a magnetic field was applied perpendicular to the plane of the dust crystal. With increasing magnetic field up to $\sim 1.1\,T$, the crystal experienced an azimuthal rotation driven by ion $\vec{v}\times\vec{B}$ drag forces. This rotation transferred momentum to the dust particles, and eventually led to melting of the crystal due to shearing.

In another set of experiments, the same chamber was used but the FTO glass plate was replaced with an interwoven copper mesh. In results first reported by Thomas et al.\cite{thomas_observations_2015}\textsuperscript{,}\cite{thomas_quasi-discrete_2015} and subsequently by Hall et al.\cite{hall_methods_2018}, at lower pressures ($p \leq 50\,mTorr$) the application of a magnetic field led to the observation that the dust particles were trapped in spatially periodic locations with a 4-fold ordering aligning with the spatial structure of the mesh. These two different observations of dust behavior with the application of a magnetic field but over a range of pressures and different boundary conditions has, over the years, led to speculation whether exists some underlying processes that could govern the spatial properties of this system.

The goal of this work was to recreate observations from these two experiments by conducting a continuous magnetic transition experiment between a hexagonal Coulomb crystal where the particles self-order with 6-fold symmetry in an unmagnetized plasma; to a state where the particles show imposed ordering with 4-fold symmetry following the presence of a conducting mesh placed above the dust in a strongly magnetized plasma. Through this experiment, some inferences are drawn about the effect that magnetic fields have on - (a) dust flow and ordering, and (b) potential structures in low-temperature plasmas. 

In section \ref{section:Experiment setup}, the experimental setup and the typical plasma parameter space are described. In section \ref{subsection:Experiment 1}, results of the magnetic transition experiment under a mesh are described. A hypothesis for the mechanism behind the phenomenon is presented and to support this hypothesis a second set of experiments in subsection \ref{subsection:Experiment 2} is conducted where the non-uniformity in boundary condition imposed by the conducting mesh is "shorted out" by placing a conducting glass on top of it. In section \ref{section:Discussions}, the physics implications of the results are discussed and in particular how these phenomena relate to ion magnetization while drawing parallels with other intriguing phenomena seen in MDPX as a result of ion magnetization.

\section{Experiment setup}\label{section:Experiment setup}

\begin{figure*}
\includegraphics[scale=0.5]{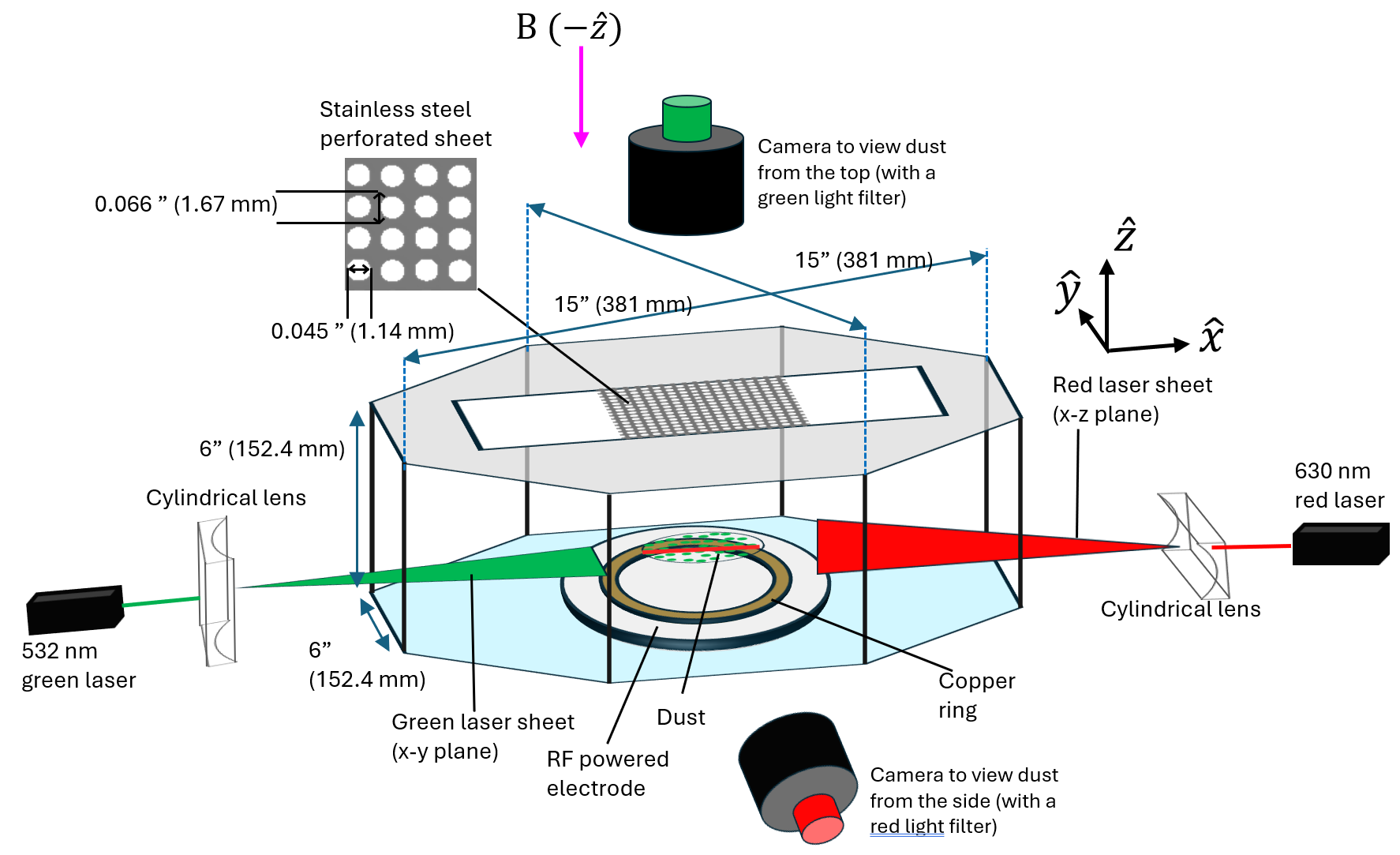}
\caption{\label{fig:Experiment1} Experiment schematic: Dusty plasma is created in an octagonal vacuum chamber and the dust is viewed using laser sheet - camera sets. Schematic is representative and not to scale.}
\end{figure*}

\begin{figure*}
\includegraphics[scale=0.42]{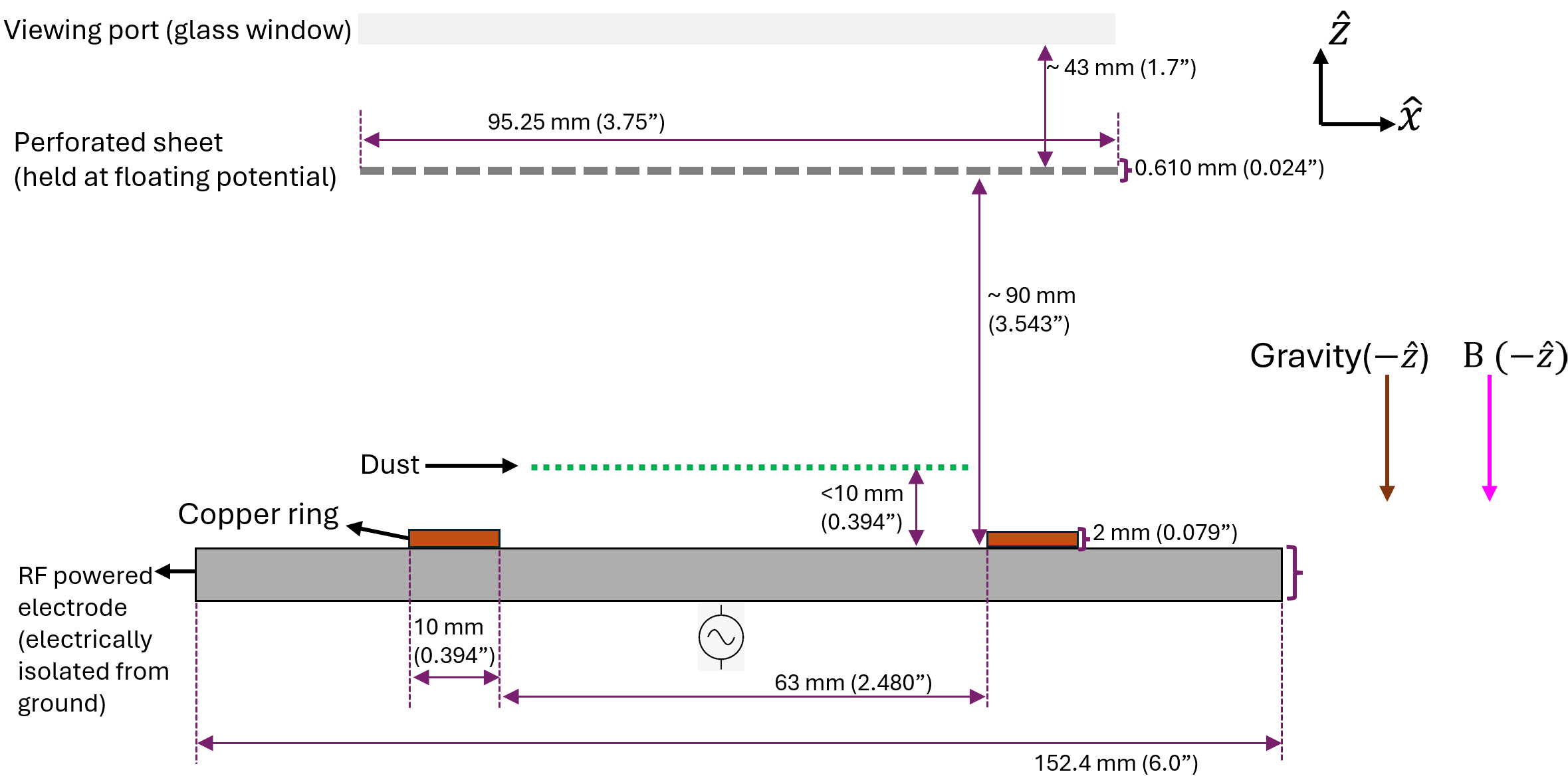}
\caption{\label{fig:Electrode dimensions} A sectional schematic of the electrode(s) and dust along with the relevant dimensions. Schematic is representative and not to scale.}
\end{figure*}

\begin{figure}
\includegraphics[scale=0.4]{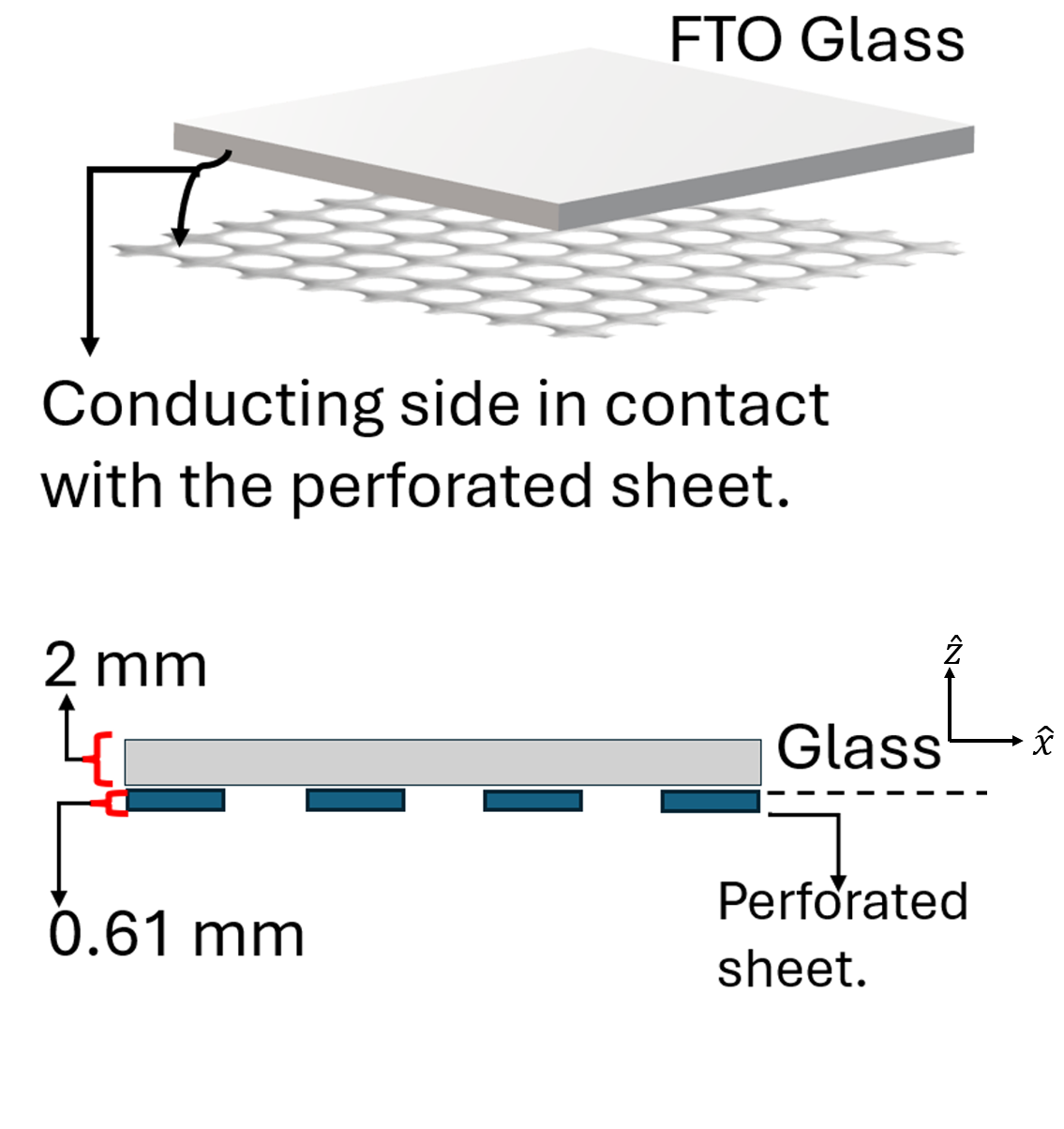}
\caption{\label{fig:FTO schematic} A schematic showing how the FTO conducting glass is placed on the perforated sheet. Schematic is representative and not to scale.}
\end{figure}

A schematic for the experiment is shown in figures \ref{fig:Experiment1} and \ref{fig:Electrode dimensions}. A capacitively coupled plasma is created by powering the lower electrode with an RF generator and auto-tuning matching network. A stainless steel perforated sheet with circular holes spaced out periodically is placed above the dust, electrically isolated from the chamber walls which are grounded. For the results described in section \ref{subsection:Experiment 2}, a conducting Fluorine doped Tin Oxide glass is placed on top of the perforated sheet as shown in figure \ref{fig:FTO schematic}.

The MDPX device consists of two main subsystems:  the magnetic field system and the plasma chamber. Magnetic fields in MDPX are produced by four superconducting coils contained in a cryostat. The coils are designed to produce a uniform  magnetic field (i.e., with less than a 1\% spatial variation) of up to 4 T in the chamber when all four coils are energized with the same current. Extensive descriptions of the MDPX coil system design are given in \cite{jr_magnetized_2015}\textsuperscript{,} \cite{thomas_magnetized_2012}. The second main subsystem of the MDPX device is the plasma chamber. This is a modular octagonal vacuum chamber of dimensions $15\,in\times15\,in\times6\,in$ ($381\,mm \times 381\,mm \times 152.4\,mm$). This chamber has glass windows on some of the eight sides as well as a glass window on the top to set up a laser sheet-camera diagnostic to view the particles. The octagon containing the electrode and mesh is placed in the center of the MDPX device. A schematic drawing of the chamber is given in figure \ref{fig:Experiment1}.  A more detailed drawing of the electrode configuration for this experiment is given in figure \ref{fig:Electrode dimensions}.

Monodisperse $7.75\,\pm\,0.29\,\mu m$ silica particles were used in these experiments as the dust particles. Dust is injected into the system by shaking a dust container with a hole which can be pushed in above the electrode when adding dust and can be pulled away so as to not interfere with the experimental region of interest above the electrode. A copper ring is placed in the center of the electrode to provide enhanced radial confinement to the dust. An upward electric force from the sheath above the electrode balances out gravity, allowing the dust to float in the $x-y$ plane, as shown in Figures \ref{fig:Experiment1} and \ref{fig:Electrode dimensions}. Laser sheets are made by allowing laser light to pass through a cylindrical lens to illuminate the dust. Two lasers are used - a $168\,mW$, $532\,nm$ green laser and a $730\,mW$, $630\,nm$ red laser. A green laser sheet illuminates the dust in the $x-y$ plane and the red laser sheet looks at a cross section of the dust in the $x-z$ plane. Two Ximea cameras with adjustable zoom lenses are placed on top and side to look at the $x-y$ and $x-z$ dynamics. Both cameras have bandpass wavelength filters to restrict plasma glow from entering the camera and capture dust motion parallel and perpendicular to the magnetic field independently of each other. Both cameras record movies at 100 frames per second. The top and side camera resolutions are approximately 40 and 60 microns per pixel respectively. The videos were processed using FIJI\cite{schindelin_fiji_2012} and particles were identified using TrackPy\cite{crocker_methods_1996}\textsuperscript{,}\cite{allan_soft-mattertrackpy_2025}.
\section{Results} \label{Section:Results}
In these experiments, the well known hexagonal Coulomb crystal is made at magnetic field $B = 0\,T$ as the initial condition. Subsequently, the magnetic field is varied to look at the different phases of spatial dust ordering as a function of magnetization of the plasma.

The spatial ordering, structure and stability of dust phases are quantitatively studied using the pair correlation function defined as:

\begin{equation}
    g(r) = \frac{Area}{2 \pi r N^2}\sum_{i}^{N} \sum_{j\neq i}^{N} \delta(r - r_j),
\end{equation}
where $Area$ is the area occupied by the crystal, N is the number of particles, and $r_j$ is the location of the individual particles. The pair correlation function effectively describes how the particle density varies with distance from any reference particle in the system. In this paper, a time-averaged $g(r)$ is reported (i.e., $g(r)$ is calculated for every frame and an average is constructed for every $r$). As shown in some of the structure factors reported in Figures \ref{fig:B_transition_1} and \ref{fig:B_transition_FTO}, the error bars at various values of $r$ in the structure factors indicate the variations in structure over time. In addition, the global maximum distance $\Delta_1$ is emphasized in these structure factors. $\Delta_1$ represents the most probable inter-particle separation, an important quantity that will be used in Section \ref{section:Discussions} to compare against the mesh spacing distance and quantify a transition in spatial ordering of dust that is observed in these experiments.

The structure and dynamical behavior of the dust at various magnetic fields can be qualitatively understood through image stacks. These images are generated by simply plotting the positions of all identified particles at all times in a single image. Effectively, these images tell us the region of space mapped out by the dust and whether some regions are more or less accessed than others.

This can be seen in Fig. \ref{fig:B_transition_1}, which shows examples of stacked images of the dust particles.  In Fig. \ref{fig:B_transition_1}(a), $B = 0 T$, the stacked (N) images of the plasma crystal reveal the spatial stability of the self-ordered crystal. By contrast, in Fig. \ref{fig:B_transition_1}(e) at $B = 0.65 T$, the stacked (N) images reveal the motion of the dust particles has formed a square-like pattern.

\subsection{Experiment I - phase transition of dust under a mesh}\label{subsection:Experiment 1}
\begin{figure*}[htbp]
	\centering
	\includegraphics[scale=0.45]{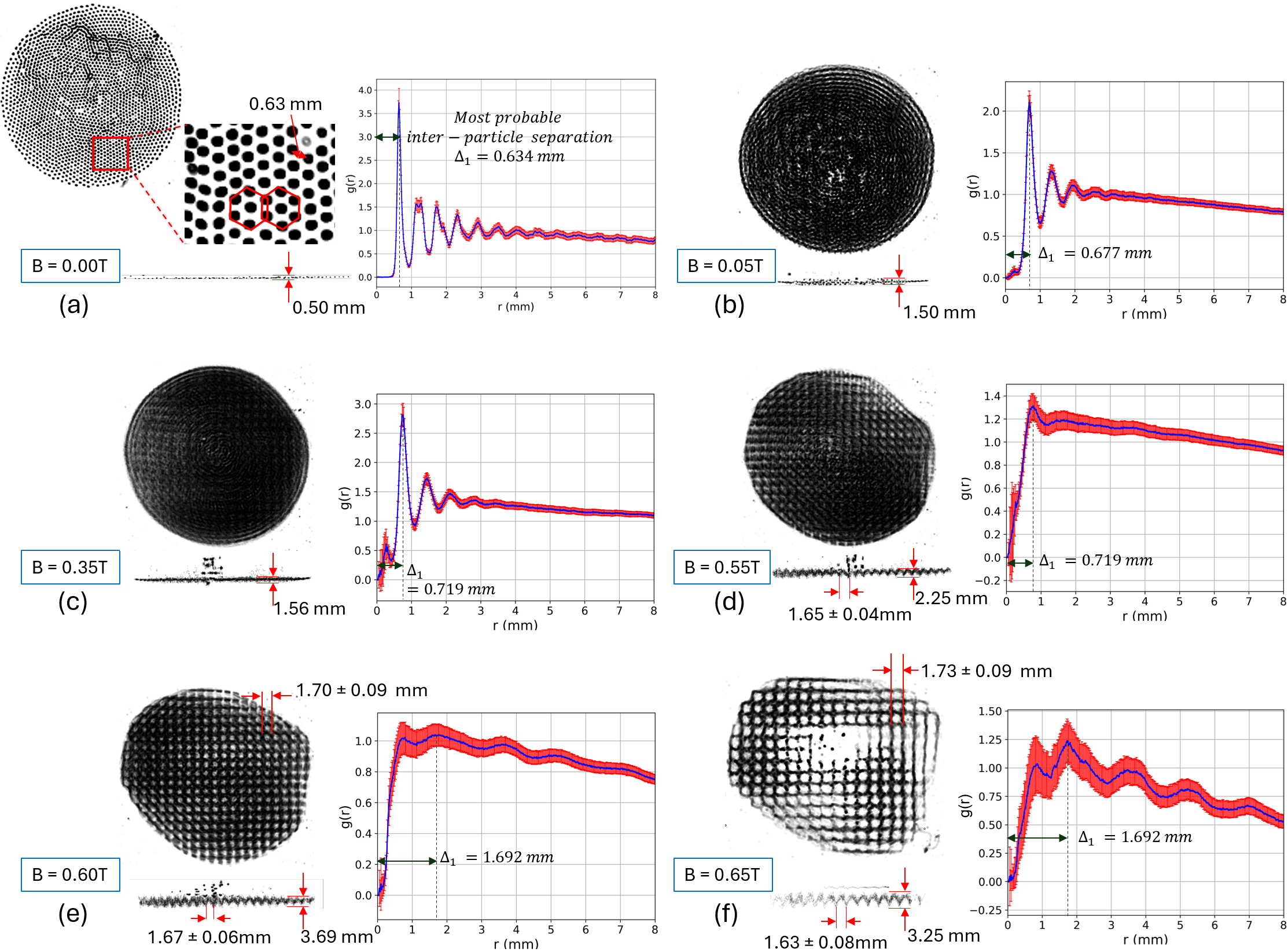}
	\caption{Dust behavior under the mesh at various magnetic fields. Each subpanel contains 500 image stacks (5 seconds) from the top ($x-y$) view and side ($x-z$) view. Next to the image stacks the time-averaged pair correlation functions are plotted. Error bars in the structure factors indicate small variations in time and a stability in the structure in time. $\Delta_1$ is the global maximum distance which represents the most probable inter-particle separation. (a) a two-dimensional hexagonal crystal with strong long-range ordering at $B=0\,T$; (b),(c) a rotating dust cloud with prominent motion parallel to the magnetic, with reduced long-range dust-dust correlations; (d) a state where there are weak signs of imposed ordering in the image stacks and little to no dust-dust correlations; (e),(f) imposed ordered dust phase where the image stacks as well as the structure factors show ordering that matches the mesh geometry.}
	\label{fig:B_transition_1}
\end{figure*}

The crystal is made in an unmagnetized plasma by introducing dust particles into the plasma just above the RF electrode at $\sim 2W$ of incident RF power. To ensure the formation of a monolayer of dust particles, the applied RF power is adjusted slightly to remove any out-of-plane particles.  An example of the initial plasma crystal state is shown in Fig. \ref{fig:B_transition_1} (a), the B = 0 T case.\par
\begin{figure*}[htbp]
	\centering
	\includegraphics[scale=0.43]{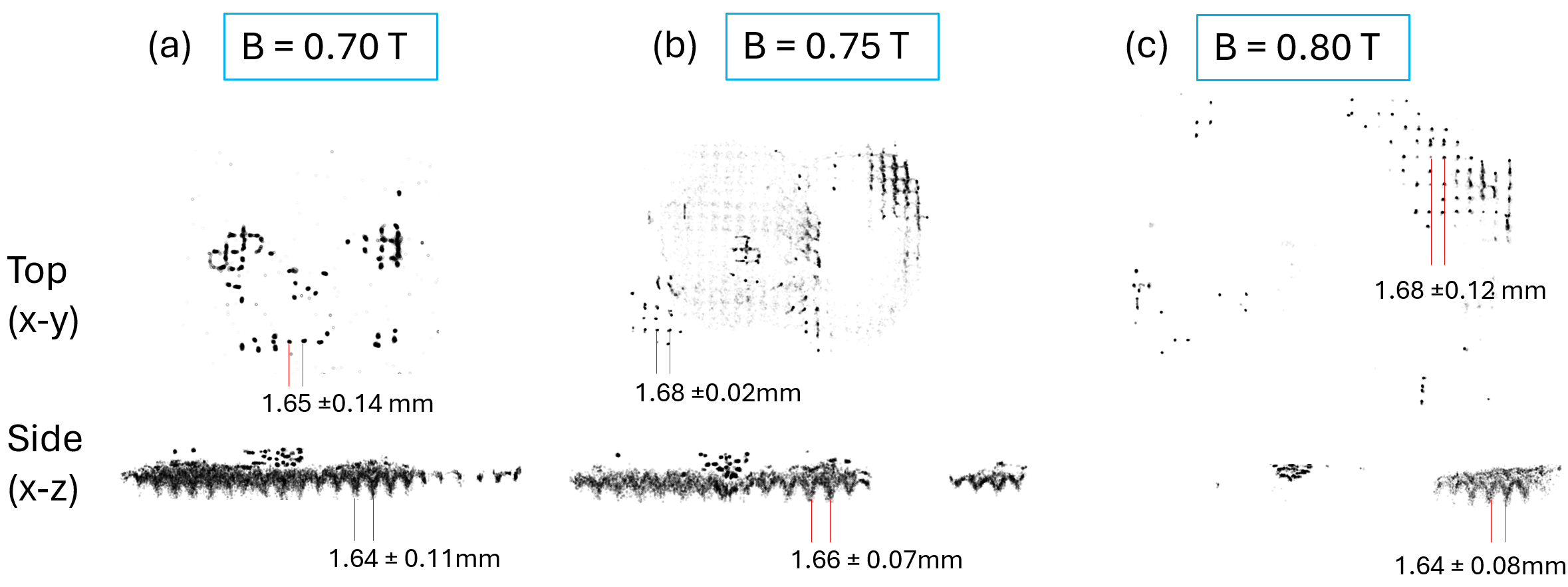}
	\caption{Image stacks of dust at magnetic fields $B>0.65\,T$. After significant particle loss for $B=0.65\,T$, more dust is added into the system and some of the particles get trapped for a long time in a 4-fold symmetry (as seen from the top image stacks), showing imposed ordering. The rest of the short lived cloud also exhibits periodicity along $\hat{x}$ that matches the mesh spacing.}
	\label{fig:B_transition_2}
\end{figure*}

After the formation of the plasma crystal at $B = 0\,T$, the magnetic field was increased in steps of $0.05\,T$ (500 Gauss) upto $B = 0.8\,T$.  For the experiments reported here, measurements are performed at a neutral pressure of $58\,mTorr$ ($7.73\,Pa$).  The experimental observations of the spatial ordering of the dust particles under the mesh electrode as a function of the magnetic field are discussed in Figs. \ref{fig:B_transition_1} and \ref{fig:B_transition_2}. \newline Fig. \ref{fig:B_transition_1} shows a comprehensive summary of the spatial structure of the dust particles at a fixed pressure ($p = 58\,mTorr$ or $7.73\,Pa$) over a range of magnetic fields from $B =0\,T$ to $B = 0.65\,T$.  Each subpanel represents data at a specific magnetic field. What is shown to the left of each subpanel is an image stack of 500 images from the top ($x-y$) and side ($x-z$) views of the dust particles. The time-averaged structure factor, $g(r)$, is shown to the right of each subpanel.

At $B=0$, a hexagonal crystal was prepared as the initial condition for the magnetic transition experiment. The top image stack indicates that particles were all largely localized and self ordered with a 6-fold symmetry. The side image stack indicates that the dust particles are in a single layer. Therefore, at $B=0$, we have a two-dimensional crystal. The pair correlation function or the structure factor for this contains maxima at equidistant locations, indicating the locations of the neighbors. The global maximum in the structure factor, which happens to be the first peak here (the nearest neighbor), indicates that given a reference particle, the likelihood of finding another particle is greatest at an average distance of $0.634\,mm$. The 7 to 8 discernible peaks in the structure factor indicate strong correlations up to the 7\textsuperscript{th} or 8\textsuperscript{th} nearest neighbors.

Once the initial condition was established, the magnetic field was energized and video data of the dust was collected at every $0.05\,T$ (500 Ga). For $B>0\,T$, the crystal rotates about a global axis of rotation. Multiple experimental studies pertaining to rotation of dust crystals and dust clouds in the presence of a magnetic field have been conducted\cite{konopka_rigid_2000}\textsuperscript{,}\cite{sato_dynamics_2001}\textsuperscript{,}\cite{jaiswal_effect_2017}. 

The top image stack show that for low values of $B>0\,T$ (for example at $B=0.05\,T$ and $B = 0.35\,T$ shown in Figs. \ref{fig:B_transition_1} (b) and \ref{fig:B_transition_1} (c)), the dust first rotates globally, effectively covering all of the region contained within the circumference of the crystal. The side image stack shows that there is increased motion parallel to the magnetic field and that the crystal is no longer two dimensional as was the case at $B=0\,T$. The time-averaged structure factors in \ref{fig:B_transition_1}(b) and \ref{fig:B_transition_1}(c) show that the average nearest neighbor distance from any particle has increased compared to the $B=0\,T$. Additionally, the long-range dust-dust correlations that were clear in the unmagnetized plasma have reduced (correlations up to the first 3 to 4 nearest neighbors are discernible). 

As the magnetic field was increased further, at $B=0.55\,T$ and beyond, signs of spatial periodicity resembling the mesh electrode started to appear. At $B=0.55\,T$ (Fig. \ref{fig:B_transition_1}(d)), the top image stacks show weak signs of periodicity in some locations. The side image stacks show that motion parallel to the magnetic field ($\hat{z}$) has a periodicity of $\sim 1.65\,mm$ along $\hat{x}$ that closely matches the mesh spacing of $1.67\,mm$. The thickness of this dust motion along $\hat{z}$ continues to increase compared to the lower magnetic fields. The structure factor shows that there are little to no long-range correlations in the dust, with weak correlation extending only to the first nearest neighbor. The reduction in long-range correlations could be attributed to a number of factors - shear driven melting in accordance with \cite{jaiswal_effect_2017}, plasma fluctuations, the non-uniform boundary condition imposed by the mesh, etc.

For $B>0.55\,T$, signs of 4-fold spatial ordering become clear through the image stacks as well as the structure factor. As shown in Figs. \ref{fig:B_transition_1}(e,f), at $B=0.60\,T$ and $B=0.65\,T$ there is a spatial periodicity that matches the mesh spacing as seen from the side camera. The top image stacks also show a spatial ordering closely matching the mesh geometry. From the structure factor, we see that the global maximum is at a distance of $\sim 1.692\,mm$, indicating that given any particle it is most likely that another particle is found at a distance that once again matches the mesh spacing. Following the terminology set by previous works on this subject\cite{thomas_observations_2015}\textsuperscript{,}\cite{thomas_quasi-discrete_2015}\textsuperscript{,}\cite{hall_methods_2018}, this phenomenon at higher magnetic fields where the geometry of the mesh is imposed onto the dust will henceforth be referred to as imposed ordering. As will be discussed further in Section \ref{section:Discussions}, the onset of imposed ordering in this system seems to occur between $B=0.55\,T$ and $B=0.60\,T$.

Note from the top image stacks in Figs. \ref{fig:B_transition_1}(e) and \ref{fig:B_transition_1}(f), that from B = 0.6 T to 0.65 T, there was a significant loss of dust particle confinement.  As a result, for measurements $B > 0.65 T$, additional particles were added, but overall confinement remained poor compared to lower magnetic fields.  Nonetheless, as indicated in Fig. \ref{fig:B_transition_2}, some of the remained trapped in the plasma and continued to retain a spatial ordering that reflected the mesh as well as continue the out-of-plane oscillation. Due to the very low number of particles, a structure factor was not plotted for these cases. Future experiments will explore the mechanisms for this particle loss.

Based on these results and findings from prior experimental \cite{thomas_quasi-discrete_2015} as well as numerical efforts\cite{menati_experimental_2020-1}\textsuperscript{,} \cite{thomas_modeling_2017}, it is hypothesized that the presence of the mesh gives rise to potential structures that are elongated parallel to the magnetic field, becoming stronger with increasing ion magnetization and affect the motion of the dust particles. To test this idea, another magnetic transition experiment was conducted but with a boundary condition that is more uniform compared to the one that is imposed by the mesh (as described in this subsection). This experiment and its results are presented in the next subsection.

\subsection{Experiment II - phase transition of dust under a mesh covered with an FTO glass}\label{subsection:Experiment 2}

\begin{figure*}[htbp]
	\centering
	\includegraphics[scale=0.42]{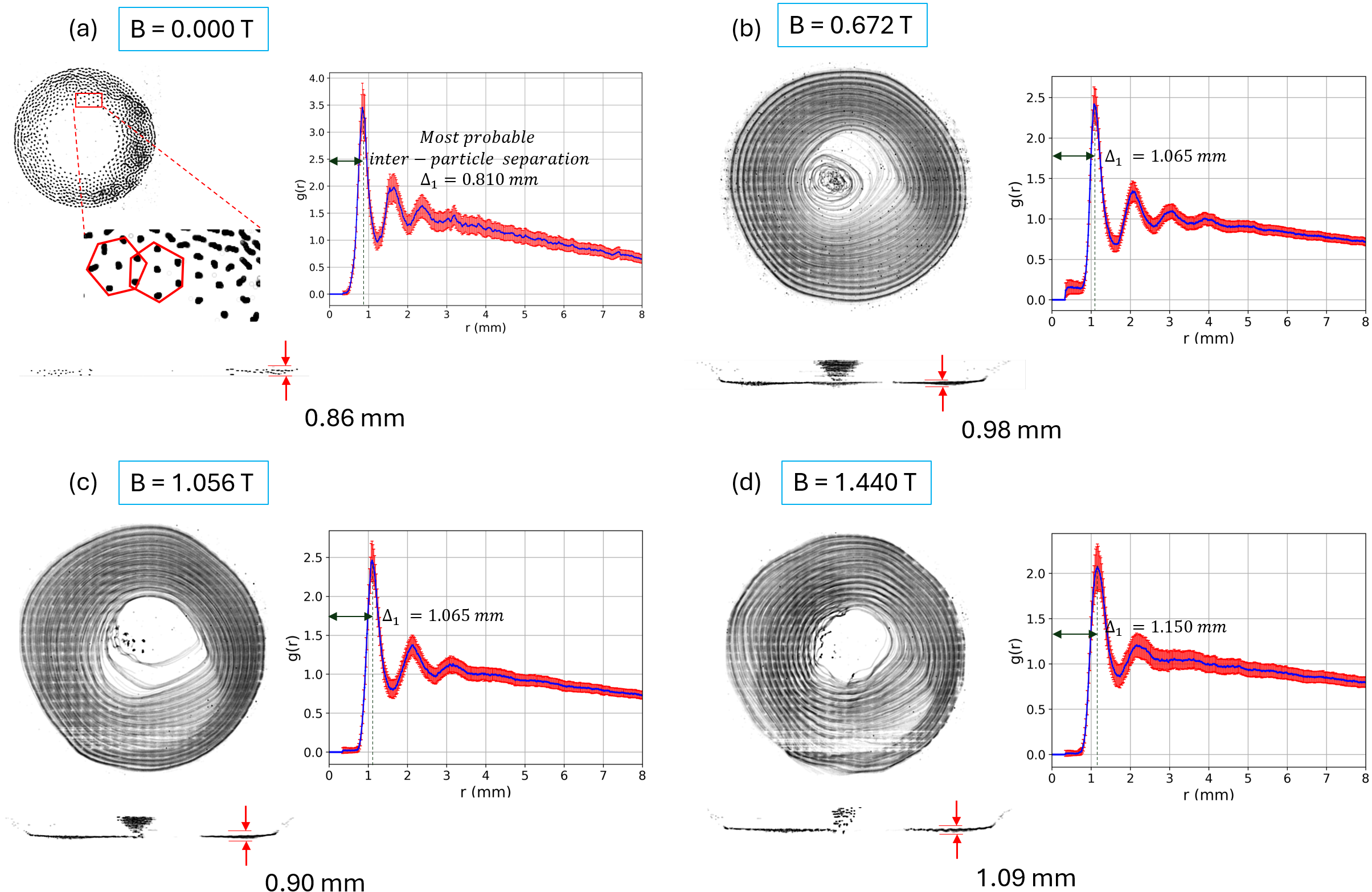}
	\caption{Dust behavior under the mesh covered with an FTO glass at various magnetic fields. Each subpanel contains 500 image stacks (5 seconds) from the top ($x-y$) view and side ($x-z$) view. Next to the image stacks the time-averaged pair correlation functions are plotted. Error bars in the structure factors indicate small variations in time and a stability in the structure in time. $\Delta_1$ is the global maximum distance which represents the most probable inter-particle separation. (a) dust crystal in an unmagnetized plasma where dust particles have 6-fold symmetry and the average nearest neighbor distance is $0.810\,mm$; (b),(c),(d) rotating dust phases with limited motion parallel to the magnetic field and weak signs of periodicity seen from top image stacks. Structure factors show no spatial ordering that matches the mesh spacing of $1.67\,mm$ even at magnetic fields much higher than that at which imposed ordering was seen in experiment 1 from Section \ref{subsection:Experiment 1}. Correlations up to the first 2-4 nearest neighbors is maintained up to $1.44\,T$.}
	\label{fig:B_transition_FTO}
\end{figure*}

In the experiment that will be discussed in this subsection, the non-uniformity inherent to the perforated sheet is reduced by placing a piece of Fluorine-doped Tin-Oxide (FTO) glass on top of it (see Fig. \ref{fig:FTO schematic}). An FTO glass is a transparent glass which is also conducting. Therefore, placing it on top of the perforated sheet effectively makes it behave like an equipotential boundary condition while allowing the light scattered by the dust to enter the top camera. The experiment protocol is the same as the one described in the previous subsection and the results are shown in Fig. \ref{fig:B_transition_FTO}. For these experiments, the incident RF power to the lower electrode is $\sim 4\,W$. The background pressure is $58\,mTorr$ ($7.73\,Pa$), same as the first set of experiments. As can be seen in the figure, there is a continuous circulation of the particles with increasing magnetic field.  This is broadly consistent with the observations reported in Jaiswal, et al. \cite{jaiswal_effect_2017}.

The dust forms a roughly two-dimensional hexagonal crystal at $B=0\,T$, showing good correlation up to the first 3-4 nearest neighbors. For $B>0\,T$ the dust starts to rotate. When compared to the dust behavior under the mesh at $B=0.65\,T$ shown in figure \ref{fig:B_transition_1} (f), there is limited evidence of imposed ordering at the comparable magnetic fields as seen from the stacked images of the top view of figure \ref{fig:B_transition_FTO}. Compared to experiments without the FTO (Fig. \ref{fig:B_transition_1}), where the peaks in the pair correlation function correspond to the mesh spacing at $B\,>\,0.6T$, in this experiment, the peaks in the pair correlation function remain independent of the mesh spacing, extending up to $B\,>\,1.4T$.

The other difference to note is that there was little loss of dust particles at higher magnetic fields when compared to the previous experiment where there was significant dust loss beyond $B=0.6\,T$ in the imposed ordering regime.

The observations from this experiment indicate that by making the boundary conditions more uniform, any electric potential structures formed from the mesh are not strong enough to affect the structure of the dust particles.

With insights from both of these experiments, the next section looks at an attempt to quantify the transition experiments and identify a critical value of magnetic field at which the onset of imposed ordering can be identified and how the physics of imposed ordering may be related to ion-magnetization.

\section{Discussion and conclusions}\label{section:Discussions}
\begin{figure}[htbp]
\includegraphics[scale=0.58]{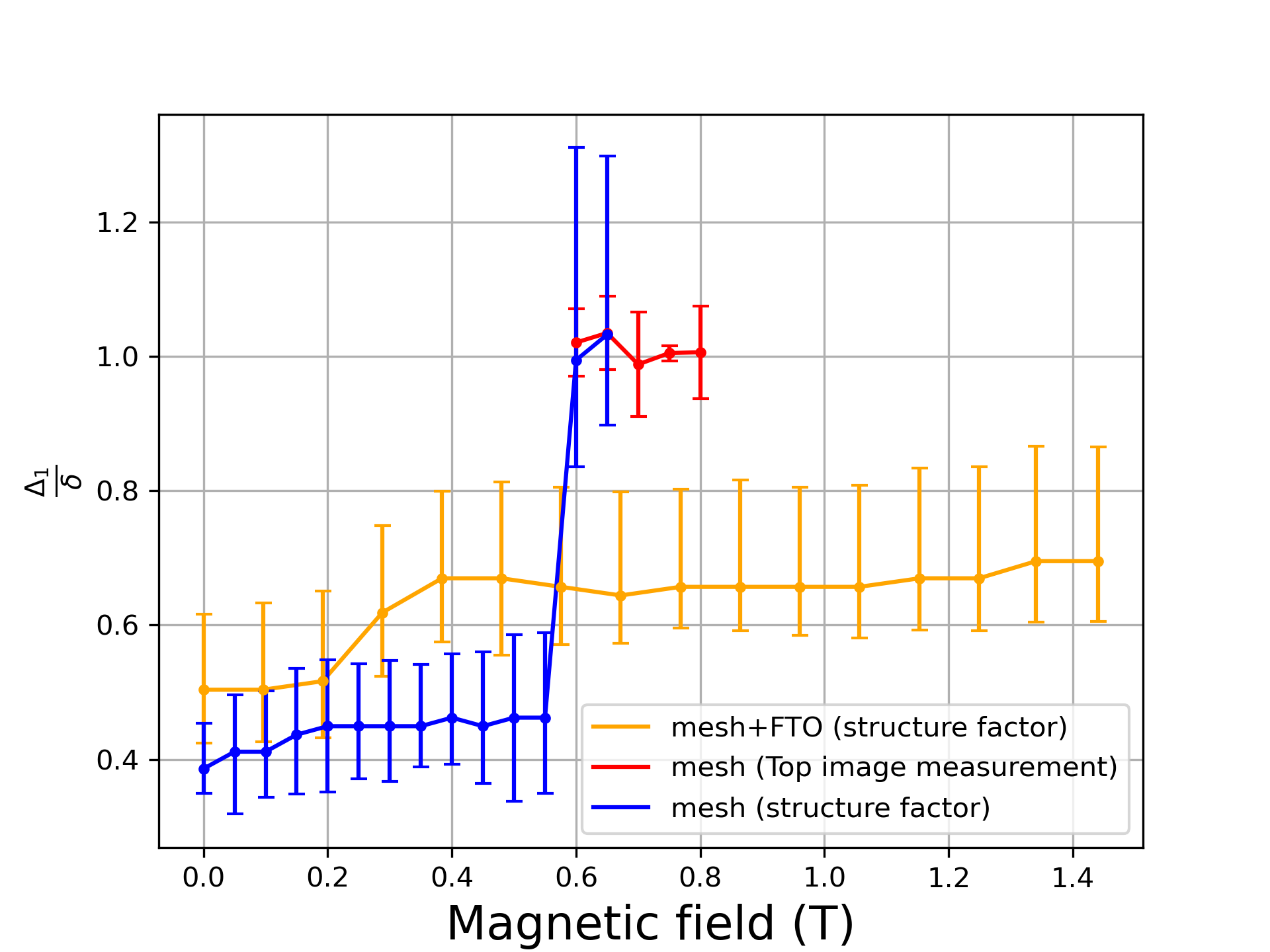}
\caption{\label{fig:Spacing_v_B} The ratio of average particle spacing distance to that of the mesh spacing $\delta = 1.67mm$ as a function of magnetic field for the transition experiments described in this work. For the continuous transition under the mesh in experiment 1 ($B \leq 0.65\,T$) and the continuous transition under a mesh covered by an FTO in experiment 2, $\Delta_1$ is the most probable inter-particle separation (i.e. the global maximum location - $\Delta_1$ from the structure factor). For these, the denoted error bars are obtained by calculating the widths of the global maximum peaks from the structure factor. For experiment 1, since no structure factor was plotted for $B>0.65\,T$, additional points are also shown where $\Delta_1$ is simply the particle spacing obtained by visual inspection of the top image stacks. It is crucial to note that spatial ordering for $B>0.65\,T$ can be obtained from side image stacks as well, which yield similar numbers as the top image stacks.}
\end{figure}

Bringing together the results from the two transition experiments discussed in Sections \ref{subsection:Experiment 1} and \ref{subsection:Experiment 2}, Fig. \ref{fig:Spacing_v_B} shows spatial ordering as a function of magnetic field. Spatial ordering of dust is quantified relative to mesh spacing as $\Delta_1/\delta$ - the ratio of the most probable inter-particle spacing to that of mesh spacing ($\delta = 1.67\,mm$). For the continuous transition under the mesh in experiment 1 ($B \leq 0.65\,T$) and the continuous transition under a mesh covered by an FTO in experiment 2, $\Delta_1$ is simply the distance from any particle at which another particle is most likely to be found (i.e. the global maximum location from the structure factor). For experiment 1, since no structure factor was plotted for $B>0.65\,T$, additional points are also shown on this plot where $\Delta_1$ is simply the particle spacing obtained by measuring the average dust-dust spacing from the top image stacks. Fig. \ref{fig:Spacing_v_B} shows that for experiment 1, there exists a sharp transition between $B=0.55\,T$ and $0.6\,T$ indicating the onset of imposed ordering ($\Delta_1/\delta\sim1$). When the mesh is covered with an FTO, no such sharp transition is seen even up to a much stronger magnetic field of $B=1.44\,T$ at the same pressures. This indicates that covering the mesh with an FTO makes the boundary equipotential and more uniform, which suppresses the formation of elongated potential structures enough to prevent or at least delay the onset of imposed ordering.

In order to provide further physical context for these results and compare the results presented in this work with prior observations of imposed ordering (conducted at different pressures and magnetic fields), it is instructive to look at ion magnetization framework\cite{hall_methods_2018}\textsuperscript{,}\cite{williams_experimental_2022}.  Here, the ion magnetization is described using a modified ion Hall parameter which is ratio of the ion gyro-circumference to the ion-neutral mean free path: 

\begin{equation}
    H_{ion} = \frac{2 \pi \rho_{gyro}}{\lambda_{i n}} = \frac{e B}{2 \sigma_{i n} p} \sqrt{\frac{k_b T_i}{8 \pi m_{i}}},
\end{equation}
where $\rho_{gyro}$ is the ion gyro-radius, $\lambda_{i n}$ is the ion-neutral collision mean free path, $e$ is the elementary charge, $\sigma_{i n}$ is the ion-neutral collision cross-section, $p$ is the neutral pressure, $k_b$ is the Boltzmann constant, $T_i$ is the ion temperature, and $m_i$ is mass of an Argon ion. 

\begin{figure}[htbp]
\includegraphics[scale=0.54]{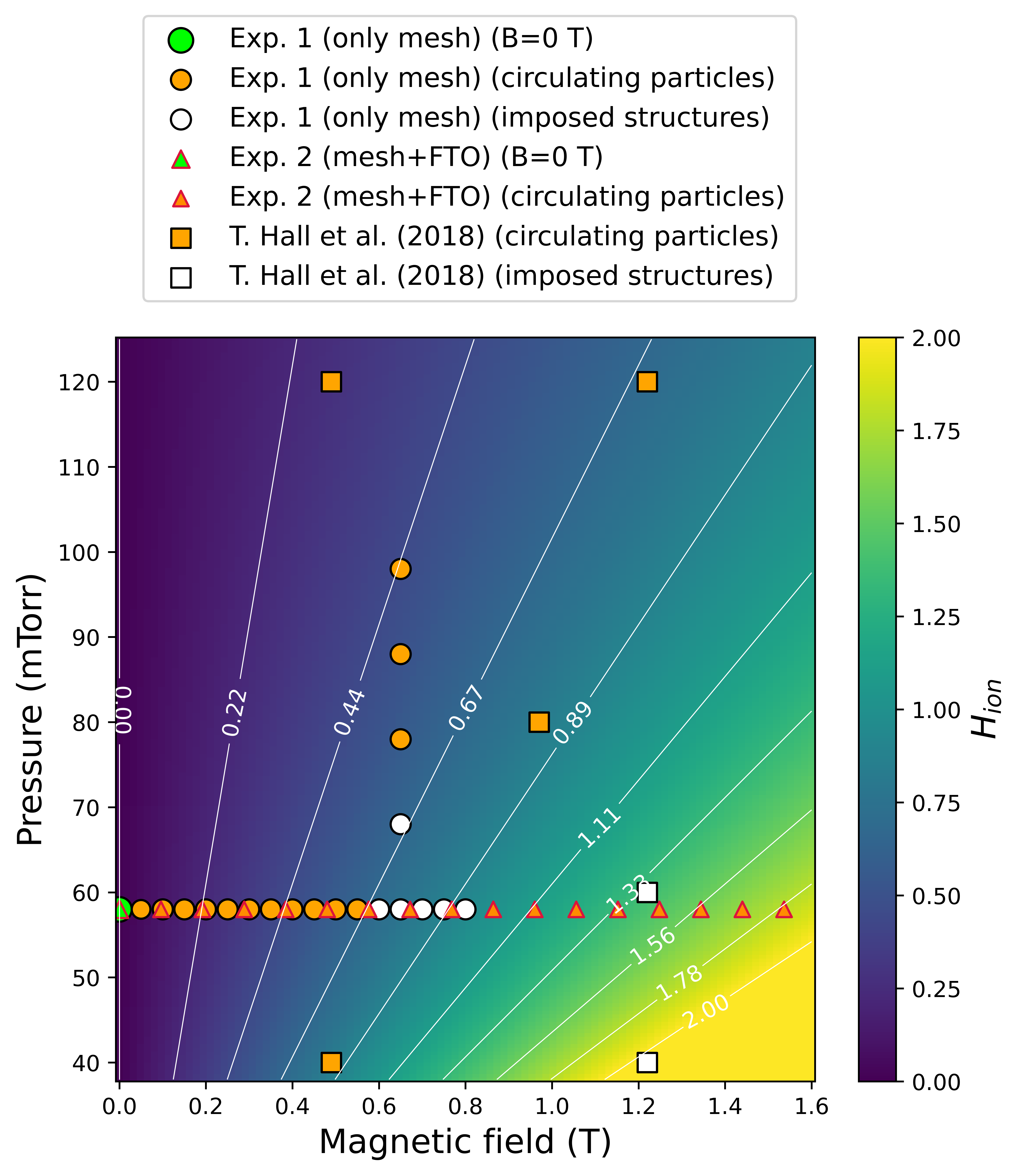}
\caption{\label{fig:Parameter_space} The magnetic field - neutral pressure parameter space consisting of the various data points from experiments that are described in this work. The points are plotted against the color-coded backdrop of the ion-Hall parameter. Imposed ordering for the experiments reported in this paper occurs for $H_{ion}>0.67$. Ion-neutral collision cross section $\sigma_{in} = 4.4\times10^{-19}\,m^2$ obtained from \cite{menati_experimental_2020}\textsuperscript{,} \cite{pitchford_lxcat_2017}\textsuperscript{,} \cite{pancheshnyi_lxcat_2012}\textsuperscript{,} \cite{carbone_data_2021}. It is assumed that ions are singly charged with a temperature $T_i = 300\,K$.}
\end{figure}

Fig. \ref{fig:Parameter_space} shows the magnetic field - pressure parameter space which has been populated with data points representing the experiments described in Sections \ref{subsection:Experiment 1} and \ref{subsection:Experiment 2} along with the experiment results reported by Hall et al. \cite{hall_methods_2018}. There are also additional data points that are part of a pressure scan conducted at $B = 0.65\,T$ for the magnetic transition under a mesh (i.e. the same boundary conditions that are a part of experiment 1 described in Section \ref{subsection:Experiment 1}). There are several pieces of information in Fig. \ref{fig:Parameter_space}. The experiment data points have been plotted in the backdrop of the ion-Hall parameter. White contour lines are also shown defining lines of constant $H_{ion}$. Circles represent the experiments presented in Section \ref{subsection:Experiment 1}, triangles represent the experiment in Section \ref{subsection:Experiment 2}, and  squares represent parameters of the imposed ordered experiment reported in Hall et al. \cite{hall_methods_2018}. The green-lime points represent dust phases in an unmagnetized plasma ($B=0\,T$). Orange shaded points represent flowing, circulating phases of dust. White shaded regions represent phases with imposed, ordered structures. For instance,  considering experiment 1, the green-lime circle is $B = 0\,T$. The orange circles between $B = 0.05\,T$ and $B = 0.55\,T$ represent the flowing, circulating dust particles.  The white circles for $B > 0.6\,T$ represent dust with imposed structures.\par


In the experiments reported in Section \ref{subsection:Experiment 1}, it can be seen that imposed ordering occurs for $H_{ion} > 0.67$. This is true for the magnetic field scan at constant pressure $p=58\,mTorr$ as well as for the pressure scan conducted at constant magnetic field $B=0.65\,T$. This implies that in order to reach an imposed ordered state from a circulating dust phase in the presence of a magnetic field, one can either increase magnetic field or decrease neutral pressure. These are both experimental knobs to increase ion magnetization which scales as $\sim B/p$. While the regime of imposed ordering in experiments reported by Hall et al.\cite{hall_methods_2018} is $H_{ion} > 1.33$, the general trend in all cases is that with increasing $H_{ion}$, flowing, circulating phases of dust are first observed and eventually at higher ion magnetization, imposed ordered structures are visible. This result is encouraging because the experiments reported in this work use different discharge conditions, a different experimental chamber and different shaped mesh than the original experiments \cite{thomas_observations_2015}\textsuperscript{,}\cite{thomas_quasi-discrete_2015}\textsuperscript{,}\cite{hall_methods_2018}. This lends support to the idea that imposed ordering is a fundamental property of the dusty plasma and that dust can reveal physical thresholds related to ion magnetization.\par
In summary, the main result of this work is the first demonstration of a continuous spatial transition of a dusty plasma from a self-organized plasma crystal to an imposed structure. While this transition generally occurs with increasing magnetic field, a more general statement is that transition appears to occur with increasing ion Hall parameter; i.e., $H_{ion}\,>\,0.67$.  Consistent with previous experiments\cite{thomas_observations_2015}, applying a conducting glass plate to the mesh suppresses the formation of the imposed structure as does increasing neutral pressure. These results not only confirm that a continuous spatial phase transition is possible, but also show that the process is controllable and previously observed results can be recovered, even under different boundary conditions.  Additionally, these results also show that a new feature, the appearance of an out-of-plane oscillation, is also spatially correlated to the mesh. The continuous nature of these experiments allows for a quantitative study of the phase transition of the dust. This involves a precise identification of the various dust phases, transitions between them, and the identification of thresholds in dust phases that can be related to formation of potential structures and ion-magnetization.
\begin{acknowledgments}
Funding for this work is provided by National Science Foundation though: Plasma Physics program (NSF-2515867), EPSCoR program (OIA-2148653), and the ECLIPSE program (NSF-2308947). Funding was also provided by the U.S. Department of Energy – Office of Fusion Energy Sciences, Plasma Science Facility program (SC-0019176). The MDPX facility was developed and built through a grant from the NSF Major Research Instrumentation (NSF-MRI) program (NSF-1126067). We thank Dr. Uwe Konopka and Dr. John Goree for useful discussions.
\end{acknowledgments}
\section*{References}\label{section:References}
\nocite{*}
\bibliography{references_1}

\end{document}